\documentclass{sig-alternate}

\toappear{} 

\usepackage{ragged2e}

\usepackage[notheorems]{dgleich-math}

\usepackage{graphicx}
\usepackage{amssymb}
\usepackage{amsmath}
\usepackage{multirow}
\usepackage{rotate}
\usepackage{subfigure}
\usepackage{epstopdf}
\usepackage{tabularx}
\usepackage{booktabs}
\usepackage{paralist}
\usepackage{url}
\usepackage{amssymb}
\usepackage{graphicx}
\usepackage{comment}
\usepackage{rotate}
\usepackage{listings}

\usepackage{algorithm}
\usepackage{algorithmicx}
\usepackage{algpseudocode}
\algnotext{EndFor}
\algnotext{EndIf}
\algnotext{EndWhile}
\algnotext{EndProcedure}
\newcommand{\algrule}[1][.2pt]{\par\vskip.5\baselineskip\hrule height #1\par\vskip.5\baselineskip}

\usepackage{wrapfig}
\usepackage{xspace}
\usepackage[table,usenames,dvipsnames]{xcolor}

\definecolor{gray}{RGB}{20,20,20}
\definecolor{greencm}{RGB}{0,153,0}
\usepackage{amssymb}
\usepackage{pifont}

\algrenewcommand{\alglinenumber}[1]{\small #1}

\definecolor{gray}{RGB}{150,150,150}
\definecolor{theblue}{RGB}{0,0,180}

\usepackage{textcase}
\usepackage{soul}

\newcommand{\be}{\begin{equation}}
\newcommand{\ee}{\end{equation}}
\newcommand{\bea}{\begin{eqnarray}}
\newcommand{\eea}{\end{eqnarray}}
\newcommand{\bit}{\begin{itemize}}
\newcommand{\eit}{\end{itemize}}

\usepackage{microtype}
\definecolor{lightgray}{rgb}{0.93,0.93,0.93}

\definecolor{lightblue}{rgb}{0.5,0.90,1.0}
\definecolor{lightgreen}{rgb}{0.5,0.92,0.5}
\definecolor{lightred}{rgb}{0.98,0.5,0.5}
\definecolor{lightyellow}{rgb}{1,0.90,0.40}

\def\@beginproof#1{\rm \trivlist \item[\hskip \labelsep{\it #1.\/}]} 
\def\@endproof{\outerparskip 0pt\endtrivlist} 
 
\def\@begintheorem#1#2{\it \trivlist \item[\hskip \labelsep{\sc #1\ #2.}]} 
\def\@opargbegintheorem#1#2#3{\it \trivlist 
      \item[\hskip \labelsep{\sc #1\ #2.\ (#3)}]} 
\def\@endtheorem{\outerparskip 0pt\endtrivlist} 

 \newtheorem{@theorem}{Theorem}[section]

\newtheorem{fact}{Fact}[section]

\graphicspath{{./}{./images/}}

\newcolumntype{H}{>{\setbox0=\hbox\bgroup}c<{\egroup}@{}}

\begin{document}


\title{Parallel Maximum Clique Algorithms with Applications to Network Analysis and Storage}

\numberofauthors{4} 
\author{
Ryan A.~Rossi, David F.~Gleich, \\ Assefaw H.~Gebremedhin\\
       \affaddr{Purdue University}\\
       \affaddr{Computer Science}\\
       \email{\{rrossi,dgleich,agebreme\}@purdue.edu}
\and
Md. Mostofa Ali Patwary\\
       \affaddr{Northwestern University}\\
       \affaddr{Electrical Engineering and}\\
       \affaddr{Computer Science}\\
       \email{mpatwary@eecs.northwestern.edu}
}

\maketitle
\begin{abstract}
We propose a fast, parallel maximum clique algorithm for large sparse graphs that is designed to exploit characteristics of social and information networks. The method exhibits a roughly linear runtime scaling  over real-world networks ranging from 1000 to 100 million nodes. In a test on a social network with 1.8 billion edges, the algorithm finds the largest clique in about 20 minutes. Our method employs a branch and bound strategy with novel and aggressive pruning techniques. For instance, we use the core number of a vertex in combination with a good heuristic clique finder to efficiently remove the vast majority of the search space. In addition, we parallelize the exploration of the search tree. During the search, processes immediately communicate changes to upper and lower bounds on the size of maximum clique, which occasionally results in a super-linear speedup because vertices with large search spaces  can be pruned by other processes. We apply the algorithm to two problems: to compute temporal strong components and to compress graphs.

\end{abstract}




\section{Introduction} \label{sec:intro}
The maximum clique problem seeks to find a clique (complete subgraph) of
the largest possible size in a given graph. 
The problem is NP-hard, even to solve in an approximate sense \cite{Khot-2001-cliques}. 
As a result one is inclined to believe that exact algorithms for finding maximum 
cliques will be too slow to be practical for large network analysis applications. 
Yet, many real-world problems have features that do not elicit  worst-case
behaviors from well-designed algorithms.    

In this manuscript, we propose a fast, state-of-the-art  
\emph{parallel exact maximum clique finder}.
And enabled by its efficiency, we use the clique finder 
(i) to investigate cliques in large social and information networks, 
(ii) to study largest temporal strong components in dynamic networks, and 
(iii) to compress graphs.

We demonstrate that finding the largest clique in big social and information networks is fast (Table~\ref{tab:max-cliques-sin}).  By way of example, we can find the maximum clique in social networks with nearly two billion edges \emph{in about 20 minutes} with a 16-processor shared memory system. 
Empirically, our method is observed to have a roughly linear runtime~(Figure~\ref{fig:scaling}) for these networks, which is remarkable in light of the fact that the problem is NP-hard.
As a point of comparison, our new solver significantly outperforms a 
recent fast maximum clique finder we developed~\cite{patt2012cliques} as well as an 
off-the-shelf clique enumerator (Section~\ref{sec:performance}).  
Consequently, we expect our new algorithm to be useful for tasks such as analyzing large networks, 
evaluation of graph generators, community detection, and 
anomaly detection, where maximum cliques are needed.

In its basic form, our algorithm is a branch and bound method 
with novel and aggressive pruning strategies.   
Several key components stand out as features contributing to its efficiency
and distinguishing it from existing algorithms.

First, the algorithm begins by finding a large clique using a near linear-time heuristic;
the obtained solution is checked for optimality before the algorithm proceeds 
any further, and the algorithm is terminated if the solution is found to be optimal.
Second, we use this heuristic clique, in combination with (tight) upper bounds 
on the largest clique,  to aggressively prune the graph.
The upper bounds are computed at the level of the input graph or individual neighborhoods.
Third, we use implicit graph edits and periodic full graph updates in order to 
keep our implementation efficient.
Fourth, we parallelize the search procedure.
The parallel search is designed such that processes (workers) immediately communicate changes to upper and lower bounds on the size of maximum clique.
As a result, vertices with especially large search spaces  can be pruned by other processes,
which occasionally results in a super-linear speedup.
Finally, our framework is tunable in the sense that the graph representation, data structures, and the implementations of the algorithm can be adapted based on the properties of 
the input graph.
The algorithm is discussed in detail in Section~\ref{sec:method}, 
its performance is evaluated in Section~\ref{sec:performance},
and the bounds it makes use of in its pruning strategies are reviewed in 
Section~\ref{sec:bounds}.

One motivation for this work came from a connection between the largest temporal
strong component of a dynamic network and maximum cliques in an associated graph.
In a network when each edge represents a contact -- a phone call, an email, or 
physical proximity -- between two entities at a specific point in time, one gets an evolving network structure~\cite{ferreira2002models} where a temporal path represents a sequence of contacts that obeys time.  A temporal strong component is a set of vertices where all pairwise temporal paths exist, just like a strong component  is a set of vertices where all pairwise paths exist. 

Surprisingly, checking if an evolving network has a temporal strong component of size 
$k$ is NP-complete~\cite{nicosia:023101,bhadra2003complexity}.  For some intuition, consider the following ``wrong'' reduction from the perspective of establishing NP-hardness. A temporal strong component of size $k$ corresponds to a clique of size $k$ in a temporal reachability graph where each edge represents a temporal path between vertices.  Finding the maximum clique, then, reveals the largest temporal strong component.  At a first glance, this is no help as even approximating the largest clique in a graph is hard~\cite{Khot-2001-cliques}.  
With a fast, well-designed algorithm, however, the connection can be exploited.
We apply our maximum clique finder for this analysis and discuss properties of temporal components we find in Twitter and phone call networks in Section~\ref{sec:tscc}.

Cliques are the most dense local structure possible in a network. Previous studies found cliques useful to compress a networks~\cite{buehrer2008-webgraph}. We tackle an easier problem and use cliques to compute a compression friendly ordering that makes many edges in the graph local. We find in Section~\ref{sec:compress} that this ordering generates results that are nearly as good as existing heuristics designed specifically for that problem.

We make all our implementations and further experiments available in an online appendix:
\newline
\centerline
{\small \url{http://www.ryanrossi.com/pmc}}

\section{Cliques in social and information networks}
\label{sec:data}

\begin{figure}[t]
\centering
\includegraphics[width=0.85\linewidth]{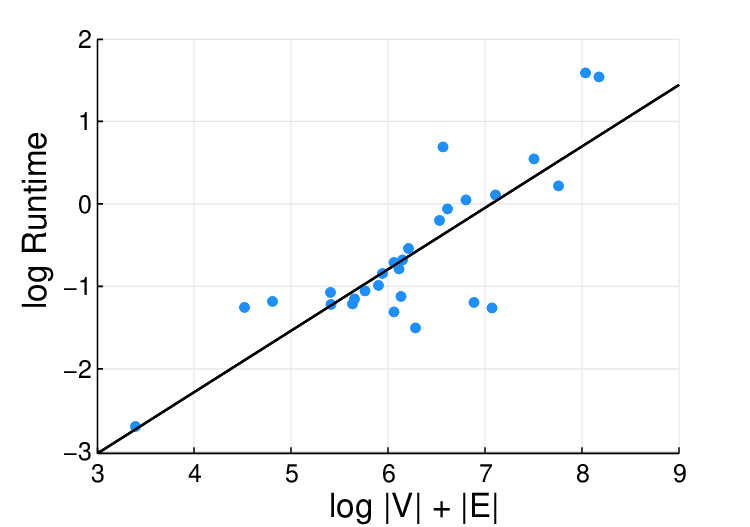}
  \caption{ The empirical runtime of our clique finder in social and
	information networks scales almost linearly with the network
	dimension.}
\label{fig:scaling}
\end{figure}

We proceed by first demonstrating how fast the algorithm finds cliques
on various social and information networks
and discussing observations we make about the cliques obtained.  
We experiment with 32 networks categorized in 8 broad classes. 
In the online appendix -- see the link above -- we present a more extensive collection of around 74 social and information networks, 
16 temporal reachability networks, and 63 dense graphs from DIMACS challenge.
Table~\ref{tab:max-cliques-sin} describes the properties of the 32 networks considered here.
It also shows the size of the largest clique in each network and states the time taken to find each clique.  
We plot the runtime pictorially in Figure~\ref{fig:scaling}, which demonstrates linear scaling 
between 1000 vertices and 100M vertices.  
We now briefly explain the data and what each clique signifies.

For all of the following networks, we discard edge weights, self-loops, and only consider the largest strongly connected component.  In contrast to the temporal components we describe later, in this section we mean the standard strong components.  If the graph is directed, we remove non-reciprocated edges. 
This strategy will identify fully-directed cliques.

\begin{table}[t!]
\caption{For each of the social and information networks studied
we find the largest clique in less than 21 minutes. 
The column $K+1$ is a core number based upper-bound,
$\tilde{\omega}$ denotes the size of the clique obtained by
the initial heuristic step,  and  $\omega$ denotes
the actual maximum clique. }
\label{tab:max-cliques-sin}
\medskip
\centering\small \scriptsize
\begin{tabularx}{\linewidth}{@{}rrXXXXHXXHH@{}}
\toprule 
& \textbf{graph} & $|V|$ & $|E|$ & $K+1$ & $\tilde{\omega}$ & $\mu$ & $\omega$ & Time \rlap{(s.)} & $\tau_{heu}$ & $\tau_{\mu}$\\ 
\midrule 
\textbf{1.} & \textsc{celegans} & 453 & 2.0k & 11 & 9 & 6 & 9 & $<$.01 & 0.33 & 1.7\\ 
 & \textsc{dmela} & 7.4k & 26k & 12 & 7 & 1 & 7 & 0.06 & 0.02 & 1.02\\ 
\midrule 
 \textbf{2.} & \textsc{mathsciet} & 333k & 821k & 25 & 25 & 1 & 25 & 0.08 & 1 & 2.16\\ 
 & \textsc{dblp} & 317k & 1.0M & 114 & 114 & 1 & 114 & 0.05 & 1 & 2.58\\ 
 & \textsc{hollywood} & 1.1M & 56M & 2209 & 2209 & 1 & 2209 & 1.69 & 1 & 7.42\\ 
\midrule 
 \textbf{3.} & \textsc{wiki}-\textsc{talk} & 92k & 361k & 59 & 14 & 2 & 15 & 0.09 & 0.22 & 1.05\\ 
\midrule 
\textbf{4.}  & \textsc{retweet} & 1.1M & 2.3M & 19 & 13 & 26 & 13 & 0.58 & 0.39 & 1.2\\ 
\midrule 
 \textbf{5.} & \textsc{whois} & 7.5k & 57k & 89 & 55 & 24 & 58 & 0.09 & 0.06 & 0.93\\ 
 & \textsc{rl}-\textsc{caida} & 191k & 608k & 33 & 17 & 31 & 17 & 0.13 & 0.35 & 1\\ 
 & \textsc{as}-\textsc{skitter} & 1.7M & 11M & 112 & 66 & 4 & 67 & 1.2 & 0.55 & 1.12\\ 
\midrule 
 \textbf{6.} & \textsc{arabic}-\textsc{2005} & 164k & 1.7M & 102 & 102 & 1 & 102 & 0.03 & 1 & 3.05\\ 
 & \textsc{wikipedia2} & 1.9M & 4.5M & 67 & 31 & 3 & 31 & 1.16 & 0.59 & 1.01\\ 
 & \textsc{it}-\textsc{2004} & 509k & 7.2M & 432 & 432 & 1 & 432 & 0.12 & 1 & 2.93\\ 
 & \textsc{uk}-\textsc{2005} & 130k & 12M & 500 & 500 & 2 & 500 & 0.06 & 1 & 6.25\\ 
\midrule 
 \textbf{7.} & \textsc{cmu} & 6.6k & 250k & 70 & 45 & 4 & 45 & 0.09 & 0.11 & 0.99\\ 
 & \textsc{mit} & 6.4k & 251k & 73 & 32 & 19 & 33 & 0.1 & 0.12 & 0.98\\ 
 & \textsc{stanford} & 12k & 568k & 92 & 51 & 768 & 51 & 0.09 & 0.26 & 2.28\\ 
 & \textsc{berkeley} & 23k & 852k & 65 & 42 & 106 & 42 & 0.16 & 0.24 & 1.03\\ 
 & \textsc{uillinois} & 31k & 1.3M & 86 & 56 & 32 & 57 & 0.18 & 0.27 & 1.13\\ 
 & \textsc{penn} & 42k & 1.4M & 63 & 43 & 32 & 44 & 0.24 & 0.27 & 1\\ 
 & \textsc{texas} & 36k & 1.6M & 82 & 49 & 34 & 51 & 0.33 & 0.23 & 0.97\\ 
 & \textsc{fb-a} & 3.1M & 24M & 75 & 23 & 35 & 25 & 6.3 & 0.41 & 1.03\\ 
 & \textsc{fb-b} & 2.9M & 21M & 64 & 23 & 196 & 24 & 5.52 & 0.38 & 1.04\\ 
 & \textsc{uci}-\textsc{uni} & 59M & 92M & 17 & 6 & 2680 & 6 & 33.86 & 0.55 & 1.33\\ 
\midrule 
 \textbf{8.} & \textsc{slashdot} & 70k & 359k & 54 & 25 & 52 & 26 & 0.06 & 0.32 & 0.95\\ 
 & \textsc{gowalla} & 197k & 950k & 52 & 29 & 2 & 29 & 0.2 & 0.24 & 1.05\\ 
 & \textsc{youtube} & 1.1M & 3.0M & 52 & 16 & 2 & 17 & 0.84 & 0.33 & 1.24\\ 
 & \textsc{flickr} & 514k & 3.2M & 310 & 45 & 82 & 58 & 5.2 & 0.04 & 1.62\\ 
 & \textsc{livejournal} & 4.0M & 28M & 214 & 214 & 1 & 214 & 2.98 & 1 & 1.49\\ 
 & \textsc{orkut} & 3.0M & 106M & 231 & 44 & 7 & 47 & 48.49 & 0.3 & 1.1\\ 
 & \textsc{twitter} & 21M & 265M & 1696 & 174 & - & 323 & 598 & 0.18 & -\\
 & \textsc{friendster} & 66M & 1.8B & 304 & 129 & 4 & 129 & 1205 & 0.45 & 1.7\\ 
\bottomrule
\end{tabularx}

\vspace{-\baselineskip}
\end{table}

\textbf{1.~Biological networks.}
We study a network where the nodes are proteins and the edges represent
protein-protein interactions (dmela~\cite{singh2008-isorank-multi}) and another where nodes are substrates and edges are metabolic reactions (celegans). 
Cliques in these networks signify biologically relevant modules.

\textbf{2.~Collaboration networks.}
These are networks in which nodes represent individuals
and edges represent scientific collaborations or movie production collaborations
(mathscinet, dblp, hollywood~\cite{boldi2004-ubicrawler}).
Large cliques in these networks are expected because they are formed when collaborations involve many participants.  

\textbf{3.~Interaction networks.}
Here, nodes represent individuals and edges represent interaction in the form of message posts (wiki-talk~\cite{leskovec2010predicting}).

\textbf{4.~Retweet networks.}
Here, nodes are Twitter users and two users are connected by an edge if  they have retweeted each other.  We collected this network ourselves. Cliques are groups of users that have all mutually retweeted each other and may represent an interest cartel or anomaly.

\textbf{5.~Technological networks.}
The nodes in these networks are 
routers (as-skitter, rl-caida, whois),
and edges are observed communications between the entities.

\textbf{6.~Web link networks.}
Here, nodes  are web-pages and edges are hyperlinks between pages 
(wikipedia, arabic-2005, it-2004, uk-2005~\cite{boldi2004-ubicrawler}). The largest clique represents the largest set of pages where full pairwise navigation is possible.

\textbf{7.~Facebook networks.}
The nodes are people and edges represent  ``Facebook friendships'' 
(CMU, MIT, Stanford, Berkeley, UIllinois, Penn, Texas~\cite{traud2011social}, fb-a, fb-b~\cite{Wilson-2009-social-networks}, uci-uni~\cite{Gjoka-2010-facebook}). 

\textbf{8.~Social networks.}
Nodes are again people and edges represent social relationships in terms of
friendship or follower (orkut, 
  LiveJournal,  
  flickr~\cite{gleich-flickr12},
  gowalla, 
  slashdot, 
  youtube\cite{leskovec2009community},
  twitter~\cite{Kwak2010-Twitter},
	friendster [Internet Archive]).

\begin{figure}[t]
\includegraphics[width=0.9\linewidth]{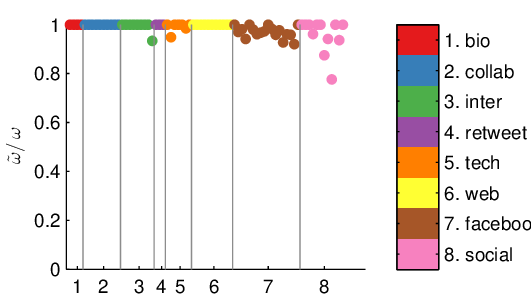}
\includegraphics[width=0.9\linewidth]{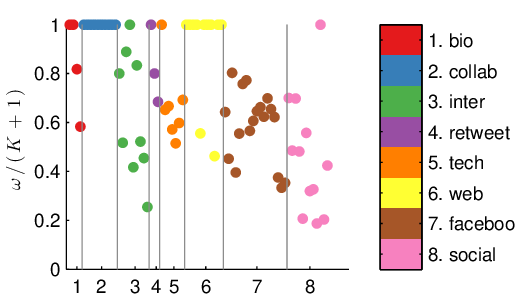}
\caption{These two figures summarize the results on all 74 networks studied in the online appendix. The top figure shows that our heuristic ($\tilde{\omega}$)  finds the largest clique ($\omega$) in biological, collaboration, and web networks in all but one case. The bottom figure identifies networks where the largest core number ($K$) tightly bounds the largest clique ($\omega$).}
\label{fig:clique-props}
\end{figure}

We summarize below our findings about cliques in these networks and the performance of our algorithm:

\begin{compactenum}[$\bullet$ \leftmargin=0em \itemindent=1em]

 \item We observe that the initial heuristic step of the algorithm  
 finds the largest clique in most cases: 17 of the 32 instances considered here,
and 52 of the 74 networks considered in the online appendix;
see Figure~\ref{fig:clique-props} for a summary.
This property helps our exact maximum clique algorithm terminate quickly.   

 \item We studied the relationship between the largest $k$-core (a notion discussed
 in Section~\ref{sec:bounds}) and the largest clique. 
 Figure~\ref{fig:clique-props} shows a summary of the results we obtained on all 74 networks.
In the collaboration and most web-link networks, we find that the largest $k$-core is a maximum clique for every graph.  The social networks, in comparison, have a much larger difference between the two, which suggests a fundamental difference in the types of networks formed via collaboration relationships versus social relationships.

\item We observe that technological networks have surprisingly large cliques.  Given that a clique represents an overly redundant set of edges, this would suggest that these maximum cliques represent over-built technology, or critical groups of nodes. 

\item We observe that for the twitter network, the nodes in the largest clique are a strange set of spam accounts and legitimate accounts with thousands of followers and following thousands. We believe that most members of this clique likely reciprocate all follower relationships. 

\end{compactenum}

\section{Bounds on  maximum clique size}
\label{sec:bounds}
As a prelude to our algorithm, we review a few easy to derive upper bounds on the size of the largest clique $\omega(G)$ in a graph $G$.  These bounds will allow us to terminate our algorithm once we have found something that hits the upper-bound or stop a local search early because there is no larger clique present.

A simple upper bound on the size of the largest clique is the maximum degree 
$\Delta(G)$ in the graph.  
Usually this is too simple to be useful.  A stronger bound can be obtained  using  $k$-cores.  
A \emph {$k$-core} in a graph $G$ is a vertex induced subgraph where all vertices have degree at least $k$~\cite{seidman1983network}.  The core number of a vertex $v$ is the largest $k$ such that $v$ is in a $k$-core. We denote it by $K(v)$.  Suppose that $G$ contains a clique of size $k$, then each vertex in the clique has degree $k-1$ and the entire graph must have a $k-1$-core.  Consequently, if $K(G)$ is the largest core number of any vertex in $G$, then $K(G)+1$ is an upper bound on the clique size.  In contrast to cliques, the core numbers of all vertices in a graph can be computed with a linear time algorithm \cite{batagelj2003m}.  

The value $K(G)$ is also known as the degeneracy of the graph.   The quantity $K(G) + 1$
is an upper-bound on the number of colors used by a greedy coloring algorithm that processes
vertices in order of decreasing core numbers -- also known as degeneracy order~\cite{Erdos-1966-chromatic-number}.  Note that the number of colors used by any greedy coloring of $G$ is also an upper-bound on the size of the largest clique because a clique of size $k$ requires $k$ colors.  Let $L(G)$ be the number of colors used by a greedy coloring algorithm that uses the degeneracy order. Then $L(G) \le K(G)+1$ and we get a potentially tighter bound on the size of the largest clique.  The bound $L(G)$ can be computed in linear time with some care on the
implementation of the greedy coloring scheme.  We summarize the bounds we have at this point:
\begin{fact}
\( \omega(G) \le L(G) \le K(G)+1 \le \Delta(G)+1. \)
\label{lm:bound1}
\end{fact}
We can further improve the bounds in Fact~\ref{lm:bound1}
by using one additional fact about a maximum clique
in a graph.  Any neighborhood graph of a vertex within the largest clique has a clique of the same size \emph{within the neighborhood graph} as well.  
The way our algorithm proceeds is by iteratively removing vertices from the graph that cannot be in the largest clique.  Let $N_R(v)$, the {\em reduced} neighborhood graph
of $v$, be the vertex-induced subgraph of $G$ corresponding to $v$ and all neighbors of $v$ that have not been removed from the graph yet.  All the bounds in Fact~\ref{lm:bound1}
apply to finding the largest clique in each of these neighborhood subgraphs.  We can therefore
state: 
\begin{fact}~\\[-2.5\baselineskip]
\begin{align}
\omega(G) & \le \max_v L(N_R(v)) \\
          & \le \max_v K(N_R(v)) + 1 \\
          & \le \max_v \Delta(N_R(v)) + 1.
\end{align}
\label{lm:bound2}
\vspace{-1.5\baselineskip}
\end{fact}
Computing the tighter bounds in Fact~\ref{lm:bound2} requires slightly more than linear work.  
For each vertex, we must form the neighborhood graph.  If we look at the union of all of these neighborhood graphs, there is a vertex in some neighborhood graph for each edge in $G$.  Thus, there are a total of $O(|E|)$ vertices in all neighborhoods.  By the same argument, there are $O(|E|+|T|)$ edges where $|T|$ is the total number of triangles in the graph. Consequently,
we can make the following statement.
\begin{fact}
The total work involved in computing the bounds in Fact~\ref{lm:bound2}
is bounded by $O(|E| + |T|)$.   
\label{lm:time}
\end{fact}

\section{Maximum Clique Algorithms}
\label{sec:method}

Given an undirected graph $G=(V,E)$, let $C_v$ denote a clique of the
largest size containing the vertex $v$.
A maximum clique in $G$ can be found by computing $C_v$ for every vertex $v$
in $V$ and then picking the largest among these.
This clearly is wasteful.
Most branch and bound type algorithms for maximum clique 
speed up the process by keeping around the size of the largest clique 
computed at any point in the course of the algorithm (\texttt{maxSoFar}) and
avoiding computation of every $C_u$, $u\in V$,
that would eventually be smaller than \texttt{maxSoFar}, a process generically
referred to as \emph{pruning}~\cite{citeulike:4058448,ostergard,Tomita-2007-branch-and-bound,patt2012cliques,jingen2013cliqueMapReduce}. 
They differ chiefly in the way the pruning is done.
The algorithm we developed in a recent work~\cite{patt2012cliques} uses a hierarchical 
pruning strategy that relies primarily on comparisons of degrees of vertices in the original input graph  with
\texttt{maxSoFar}, effectively using the weakest bound in Fact~\ref{lm:bound2}.
In comparison, our new method uses the tightest bound in Fact~\ref{lm:bound2}.
Furthermore, it is parallelized, and it contains a variety of new algorithmic and performance
optimization ingredients that result in significantly superior performance.

For reference throughout the discussion in this section, we outline our algorithm in the
psuedocodes in Algorithm~\ref{fig:heuristic} and Algorithm~\ref{fig:clique-alg} and provide
an illustrative example in Figure~\ref{fig:example}. 
In the overall procedure,  we identify the following steps as the most important: 
\begin{compactitem}[--]
\item finding a good initial clique via our heuristic,
\item using the smallest to largest ordering in the main loop; this helps ensure that 
neighborhoods of high degree vertices are as small as possible,
\item using efficient data structures for all the operations and graph updates, and
\item aggressively using $k$-core bounds and coloring bounds to remove vertices early.
\end{compactitem}

\algrenewcommand{\alglinenumber}[1]{\scriptsize #1}
\begin{algorithm}[t!]
\centering
\caption{Our greedy heuristic to find a large clique. This is used as the first step
in the exact algorithm, outlined in Algorithm~\ref{fig:clique-alg}. The input array $K$
holds core numbers of vertices.}
\label{fig:heuristic}{\small
\begin{algorithmic}[1]
\Procedure {HeuristicClique}{$G=\left (V,E\right ), K$}
\State Set $H = \{\}$, Set $\text{max} = 0$
\For{each $v \in V$ in decreasing core number order}
			\If{$v$'s core number is $\geq$ max}
					\State Let $S$ be the neighs. of $v$ with core numbers $\geq$ max
					\State Set $C = \{\}$
					\For{each vertex $u \in S$ by decreasing core num.}
							\If{$C \cup \{u\}$ is a clique} 
							    \State Add $u$ to $C$ 
							\EndIf
					\EndFor
					\If{$|C| > \text{max}$} 
					      \State Set $H = C$, Set $\text{max} = |H|$
					\EndIf
			\EndIf
\EndFor
\State \textbf{return} $H$, a large clique in $G$
\EndProcedure
\end{algorithmic}}
\end{algorithm}

\subsection{Our fast heuristic clique finder}
\label{sec:heuristic}
Our exact maximum clique finder begins by calling a fast heuristic clique finder
(see the main procedure \textsc{MaxClique} in Algorithm~\ref{fig:clique-alg}).
The goal of this initial heuristic step is to find a large clique in the graph quickly.  
The heuristic is similar to the  maximum-degree based heuristic described in \cite{patt2012cliques}, which, in exploring for a maximum clique in which a vertex $v$ participates, simply picks a vertex of the highest degree in the neighborhood of $v$.  Our heuristic search differs as we use the core numbers of each vertex to guide the search instead.  The inspiration for this change is the relationship between core numbers, the degeneracy order, and a simple $2$-approximation algorithm for the densest subgraph.

The heuristic, outlined in Algorithm~\ref{fig:heuristic}, builds a clique by searching around each vertex in the graph and greedily adding vertices from the neighborhood as long as they form a clique. The order of vertices is the degeneracy order (the input parameter $K$ contains the needed core numbers of the 
vertices).  Because the core numbers are also a lower-bound on the size of the largest clique a vertex participates in, we can efficiently prune the greedy exploration.  

As mentioned in the previous section, this heuristic step in itself finds
the \emph{largest clique} in the graph in over half of the social networks we consider.
It can therefore be used as a stand-alone procedure.
All steps in Algorithm~\ref{fig:heuristic}, except for the statements in
Lines 7--9, can be performed using work proportional to the degree of a vertex.
Those statements in turn require work proportional to the size of the subgraph
induced by the neighborhood of a vertex. The overall runtime is therefore
$O(|E| \cdot \Delta(G))$. 

\subsection{Initial pruning}
\label{sec:pruning}

After our exact algorithm finds a heuristic clique $H$ in the input graph $G$
using the core numbers of the vertices, it puts those numbers to another strategic use. 
Suppose we find a clique in $G$ of size $\tilde{\omega} = |H|$. 
Then we can eliminate all vertices with core numbers strictly less than 
$\tilde{\omega}$ from our search (Line 4 in \textsc{MaxClique}). 
This pruning operation works because a clique of size $\tilde{\omega}+1$ or larger must have vertices with core numbers at least $\tilde{\omega}$.  
In a few cases, we observed this step suffices to certify that $H$ is the maximum clique 
as we remove all of the graph. 
This happens, for instance, with the LiveJournal network.
Moreover, this pruning procedure reduces memory requirements significantly 
for most networks.

In our implementation, for this initial pruning, vertices are explicitly removed from the graph.
However, vertices removed in future pruning steps are simply marked as deleted in an index array. Future graph operations, such as neighborhood queries, check this array before returning their contents. This helps us achieve efficient implementation.

\subsection{Searching} 
\label{sec:clique-search}

After we reduce the size of the graph via the initial pruning, we then run a search strategy over all the remaining vertex neighborhoods in the graph (the \textbf{while}-loop in \textsc{MaxClique}).  The algorithm we run is similar to a standard branch and bound scheme for maximal clique enumeration~\cite{Bron-1973-all-cliques}. However, we unroll the first two levels of branching and apply our clique bounds in order to find only the largest clique.

At this point, we wish to introduce a bit of terminology.  Recall 
(from Section~\ref{sec:bounds}) that $N_R(v)$ is the reduced neighborhood 
graph of $v$ and let $d_R(v)$ denote the reduced degree of $v$.  
These sets do not contain any vertices that have been removed from the graph due to changes in the lower-bound on the clique size due to $k$-cores and any vertices whose local searches have terminated. At the risk of being overly formal, let $\tilde{\omega}$ be the current best lower-bound on the clique size, and let $X$ be a set of vertices removed via searching.  Then:
\[ N_R(v) = G( \{ v \} \cup \{ u : (u,v) \in E, K(u) \ge \tilde{\omega}, u \not \in X \}). \]


\algrenewcommand{\alglinenumber}[1]{\small #1}
\begin{figure}[t!]
\begin{center}
\begin{minipage}{0.49\textwidth}
\begin{algorithm}[H]
\centering
\caption{Our exact maximum clique algorithm. See Section~\ref{sec:parallelization}
for details about how to parallelize it.}
\label{fig:clique-alg}
{
\small
\begin{algorithmic}[1]
\Procedure {MaxClique}{$G=\left (V,E\right )$}
\State Set $K$ = \textsc{CoreNumbers}($G$) \Comment{{\footnotesize $K$ is a vertex-indexed array}}   
\State Set $H$ = \textsc{HeuristicClique}($G, K$) 
\State Remove (explicitly) vertices with $K(v) < |H|$ 
\While{$|G| > 0$}
  		\State Let $u$ be the vertex with smallest reduced degree
  		\State \textsc{InitialBranch}($u$) \Comment{{\footnotesize the routine grows $H$}}
  		\State Remove $u$ from $G$
  		\State Periodically, explicitly remove vertices from $G$
\EndWhile
\State \textbf{Return} $H$, the largest clique in $G$
\EndProcedure
\smallskip
\algrule[0.5pt]
\smallskip
\Procedure {InitialBranch}{$u$}
\State Set $P = N_R(u)$ 
\If{$|P| \le |H|$} \textbf{return} \EndIf 
\State Set $K_N$ = \textsc{CoreNumbers}(P) 
\State Set $K(P) = \max_{v \in P} K_N(v)$
\If{$K(P)+1 <  |H|$} \textbf{return} \EndIf
\State Remove any vertex with $K_N(v) <  |H|$ from $P$
\State Set $L = $ \textsc{Color($P$, $K_N$)} in degen. order \Comment{{\footnotesize $L$ is nr of colors}}        
\If{ $L \le |H|$} \textbf{return} \EndIf
\State \textsc{Branch}($\{\}$, $P$) 
\EndProcedure
\smallskip
\algrule[0.5pt]
\smallskip
\Procedure {Branch}{$C, P$}
\While{ $|P| > 0$ and $|P| + |C| > |H|$}
  	\State Select a vertex $u$ from $P$ and remove $u$ from $P$
  	\State Set $C' = C \cup \{ u \}$
  	\State Set $P' = P \cap \{ N_R(u) \}$
  	\If{ $|P'| > 0$}
    			\State Set $L = $ \textsc{Color}($P'$) in natural (any) order 
    			\If{ $|C'| + L > |H|$} \State \textsc{Branch}($C'$, $P'$) \EndIf
  	\ElsIf{$|C'| > |H|$}  \Comment{{\footnotesize $C'$ is maximal}}    
    		\State Set $H = C'$ \Comment{{\footnotesize new max clique}}
    		\State Remove any $v$ with $K(v) < |H|$ from $G$ \Comment{{\footnotesize implicitly}}
  	\EndIf
\EndWhile
\EndProcedure
\end{algorithmic}}
\end{algorithm}
\end{minipage}
\end{center}
\end{figure}

\begin{figure}[t!]
\centering
\includegraphics[width=0.8\linewidth]{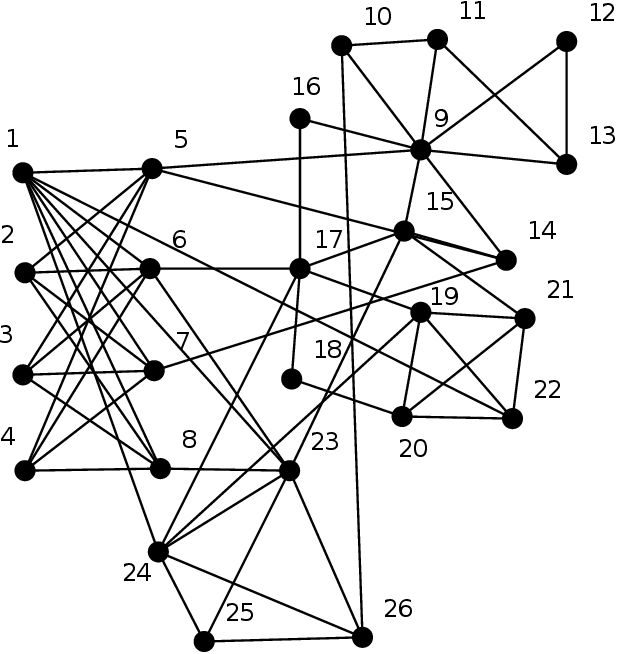}
\caption{An example that illustrates the workings of Algorithm~\ref{fig:clique-alg}.
See discussion at the end of Section~\ref{sec:clique-search}.}
\label{fig:example}
\end{figure}

We explore the remaining vertices in order of the smallest to largest reduced degree.
For each vertex, we explore its neighborhood using the function \textsc{InitialBranch}.  
When \textsc{InitialBranch} returns, we have found the largest clique involving that vertex, and so we can remove it from the graph.  Again, this is done by marking it as removed in an array.  We did, however, find it advantageous to periodically recreate the graph data structure in light of all the edits and recompute $k$-cores.  This reduces the cost of the intersection operations. In addition, we believe that this step aggregates memory access to a more compact region thereby improving caching on the processor. We do this every four seconds of wall clock time.  

The first step of \textsc{InitialBranch} is a set of tests to check 
if any of the bounds from Lemma~\ref{lm:bound2}
rule out finding a bigger clique in the neighborhood of $u$.  
The first test (Line 13) essentially corresponds to the weakest bound,
Equation (3), in Lemma~\ref{lm:bound2}.
To check against the bound given by Equation (2), 
we compute the core numbers for each vertex in the neighborhood subgraph.  If the largest core number in the neighborhood graph is no better than the current lower bound, we immediately return and add the vertex to the list of searched vertices.  
If it isn't, then we compute a greedy coloring of the subgraph using the degeneracy order. 
Using the coloring bound from Lemma~\ref{lm:bound2} (Equation (1)), we can immediately return if there is no large clique present.  If none of these checks pass, then we enter into a recursive search procedure that examines all subsets of the neighborhood in a search for cliques via the \textsc{Branch} function.

The \texttt{Branch} function maintains a subgraph $P$ and a clique $C$.  The invariant shared by these sets is that we can add any vertex from $P$ to $C$ and get a clique one vertex larger.  We pick a vertex and do this.  To be precise, we pick the vertex with the largest color (note that colors are positive integer numbers).
We then check if the clique $C'$ is maximal by testing if there is any set $P'$ that exists that satisfies the invariant. 
If it is a maximal clique, then we check against our current best clique $H$, and update it if we found a larger clique.
If it is not, then we test if it is possible that $C'$ and $P'$ have a big clique.
The biggest clique possible is $|C'| + \omega(P') \le |C'| + L(P')$, and so using the function \textsc{Color},  
we compute a new greedy coloring to get the upper bound $L(P')$. Unlike the greedy coloring in
\textsc{InitialBranch}, here we do not use the degeneracy ordering as it was not worth the extra work in our investigations.  
If $C'$ and $P'$ pass these tests, we recurse on $C'$ and $P'$.

The example in Figure~\ref{fig:example} illustrates several of the points 
we have been discussing thus far.
The core number of this graph is $4$, which yields the upper bound of $5$ on the maximum clique size. The clique detected by our heuristic is $\{1,8,23\}$; the graph has two maximum cliques: $\{19,20,21,22\}$ and 
$\{23,24,25,26\}$. Our algorithm removes vertices $10, 11, 12, 13$ and $16, 17, 18$ in the initial pruning. Subsequently, our method will explore vertex $9$ and remove it based on the maximum neighborhood core of $3$. It explores vertex $15$ next and removes it due to the neighborhood core bound. It then removes vertex $14$ due to an insufficient degree. Subsequently, it finds the clique around vertex $19$, then prunes all vertices except $1$ through $8$ due to core number bounds. Finally, it eliminates vertex $1$ due to the neighborhood core bound; all other vertices are then iteratively removed via degree bounds.

\begin{figure}[b!]
\centering
\includegraphics[width=0.85\linewidth]{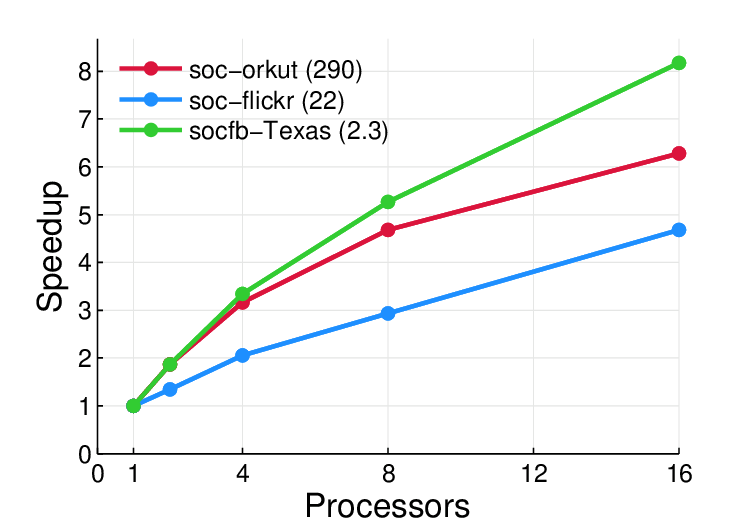}
  \caption{Speedup of our parallel maximum clique algorithm on social and information networks. Single processor runtimes in seconds are shown in parentheses.}
\label{fig:speedup-social}
\end{figure}

\subsection{Performance Optimization}

In the interest of space and  to keep the presentation simple, we have left out 
several details on
performance enhancement that we have in our implementation. 
(We make the code available online for interested readers).
To give a small example, we use an adjacency matrix structure for small graphs in order to facilitate $O(1)$ edge checks.  We use a fast $O(d)$ neighborhood set intersection procedure, and have many other optimizations throughout the code.  


\subsection{Parallelization}
\label{sec:parallelization}
We have parallelized the search procedure.  Our own implementation uses shared memory, but we describe the parallelization such that it could be used with a distributed memory architecture as well.  

The parallel constructs we use are a worker task-queue and a global broadcast channel.  
In fact, the basic algorithm remains the same.  
We compute the majority of the preprocessing work in serial with the exception of a parallel search for the clique in the initial heuristic step.  
Here, we assume that each worker has a copy of the graph and distribute vertices to workers to find the largest heuristic clique in the neighborhood.  
In serial, we reduce the graph in light of the bounds, and then re-distribute a copy of the graph to all workers.  
At this point, we view the main while loop as a task generator and farm the current vertex out to a worker to find the largest clique in that neighborhood.  
Workers cooperate by communicating improved bounds between each other whenever they find a clique and whenever they remove a vertex from the graph using the shared broadcast channel.  
When a worker receives an updated bound, we have found that it is often possible for that worker to terminate its own search at once. 
Unlike previous algorithms, the speedup from our parallel maximum clique algorithms can be \emph{super linear} since we are less dependent on the precise order of vertices explored.
In our own shared memory implementation, we avoid some of the communications by using global arrays and locked updates.


\begin{figure}[t!]
\centering
\includegraphics[width=0.85\linewidth]{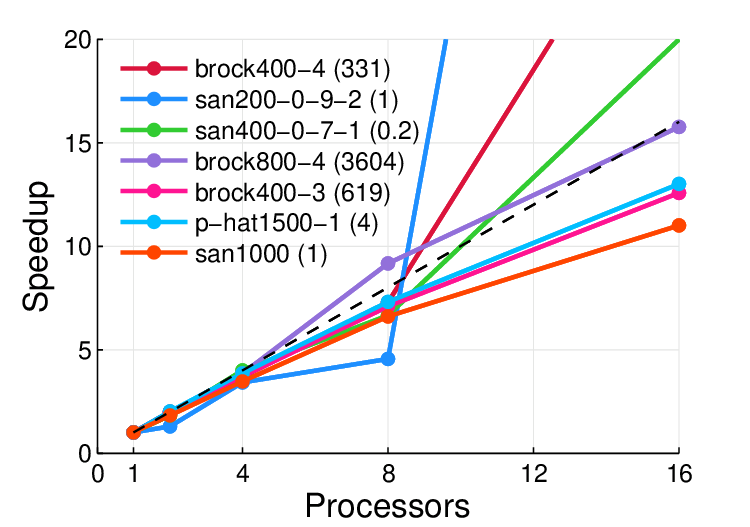}
  \caption{Speedup of our parallel maximum clique algorithm on DIMACS graphs.
Single processor runtimes in seconds are shown in parentheses. }
\label{fig:speedup-dimacs}
\end{figure}

\section{Performance Results}
\label{sec:performance}

As we illustrated in Table~\ref{tab:max-cliques-sin} and Figure~\ref{fig:scaling}, the runtime of our clique finder on social and information networks is fast, and it exhibits roughly linear scaling as we increase the problem size.  We used a two processor, Intel E5-2760 system with 16 cores and 256 GB of memory for those tests and the remaining tests.  None of the experiments came close to using all the memory. In this section we will be concerned with three questions: 
\begin{compactenum}[a)]
\item How scalable is our parallel algorithm? 
\item How does our method compare to other clique finders on social and information networks? 
\item Is the tighter upper bound that results from using neighborhood cores worth the additional expense?  
\end{compactenum}
In what follows, we will refer to our own algorithm outlined in 
Algorithm~\ref{fig:heuristic}~and~\ref{fig:clique-alg} as ``pmc'' (for parallel maximum clique).
For performance analysis purposes we will consider several versions of pmc.



For the results reported in this section, we will use problems from the 20 year old DIMACS clique challenge~\cite{Trick-1996-dimacs} to study the performance of our clique finder on an established benchmark of difficult problems.  Even the best state of the art algorithms cannot solve all of these problems. In the interest of space, we do not present individual data on them.  These graphs are all small: 45---1500 vertices.  However, they contain an enormous number of edges and triangles compared with social networks.  The number of triangles ranges between 34,000 and 520 million.  Of the 57 graphs our method was able to solve, we divide them into an easy set of 26 graphs, where our algorithm terminates in less than a second and a hard set of 32 graphs which take between one second and an hour.  

\textbf{Parallel Speedup.} In Figure~\ref{fig:speedup-social} we show the speedup obtained
as we add processes to our pmc method for three social networks. In Figure~\ref{fig:speedup-dimacs} we show speedup results of pmc for seven of the DIMACS graphs.   The runtime for both includes all the serialized preprocessing work, such as computing the core numbers initially. The figures illustrate two different behaviors. For social networks, we only get mild speedups on 16-cores for the largest problem (soc-orkut).  For the DIMACS graphs, we observe roughly linear and, sometimes, super-linear performance as we increase the number of processes.  The superlinear performance is due to quicker returns from unfruitful branches as
a result of the parallel exploration.
These results indicate that our parallelization scales well and helps reduce the runtime for difficult problems.

\begin{figure}[t]
\centering
\includegraphics[width=0.9\linewidth]{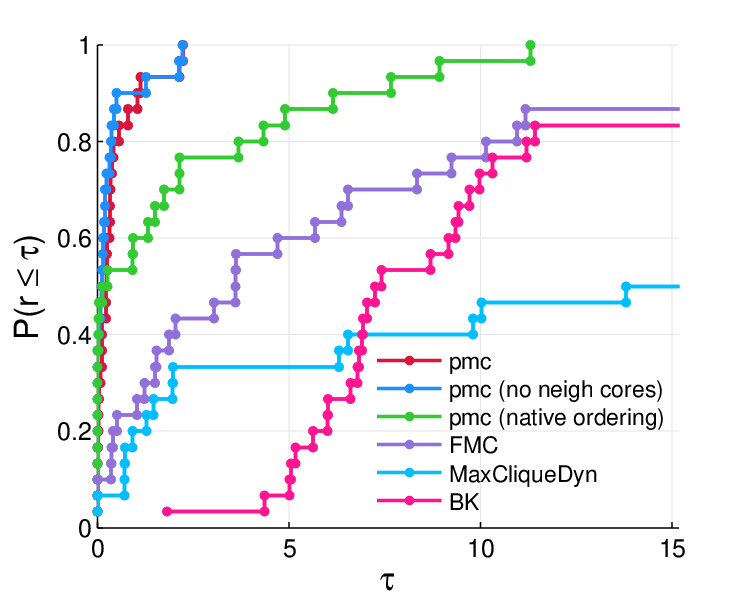}
\caption{
Comparison of a serial version of pmc and its variants against three existing 
maximum clique algorithms on 30 difficult social and information networks.}
\label{fig:pp-social}
\end{figure}

\textbf{Performance Profile Plots.} To address the two remaining questions b) and c), we use  performance profile plots to compare algorithms~\cite{dolan2002-performance-profiles}. 
Performance profile plots compare the performance of multiple algorithms on a range of problems. They are similar to ROC curves in that the best results are curves that lie towards the upper left. Suppose we have $N$ problems in total and that an algorithm solves $M$ of them within 4 times the speed of the best solver for each problem.  Then we would have a point $(\tau,p) = (\log_2 4, M/N)$.  Note that the horizontal axes reflects a speed difference factor of $2^\tau$. The fraction of problems that an algorithm cannot solve is given by the right-most point on the curve.  
In Figure~\ref{fig:pp-social}, for instance, the method labeled BK
only solves around 80\% of the problems in the test set.  

\begin{figure}[t]
\centering
   		\subfigure[DIMACS-Hard (Serial)]
    			{\hspace{-2ex}\label{fig:pp-serial}\includegraphics[width=0.7\linewidth]{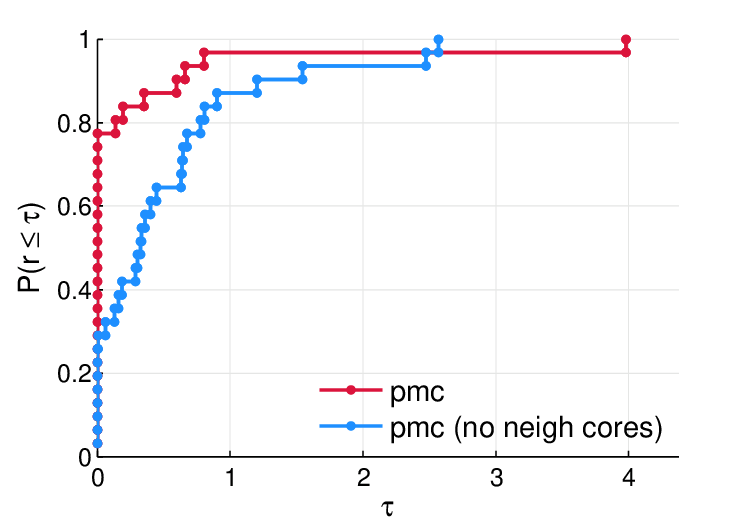}}    
			\subfigure[DIMACS-Hard (16 Threads)]
    			{\hspace{-4ex}\label{fig:pp-16}\includegraphics[width=0.7\linewidth]{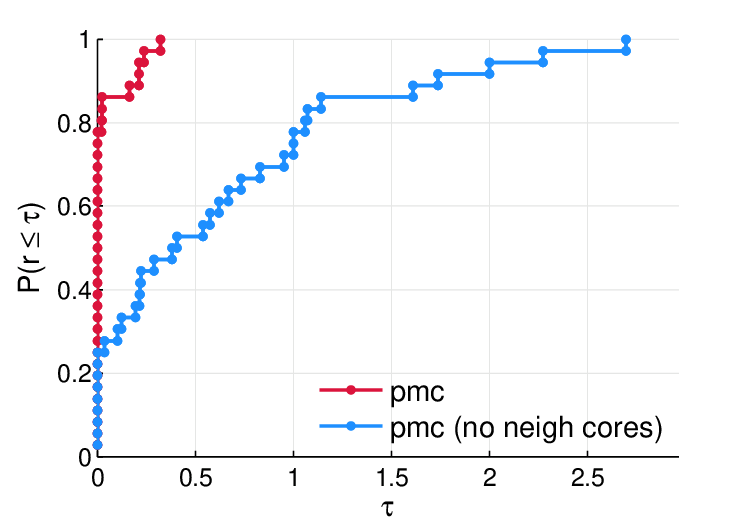}} 
  \caption{
Performance profiles comparing the performance of two versions of pmc (with and without
neighborhood cores) on DIMACS graphs.
The top figure shows comparison of serial versions of the two variants, 
the bottom figure shows similar comparison of the parallel versions.
}
\label{fig:pp-plots}
\end{figure}

\textbf{Figure~\ref{fig:pp-social}, social and information networks.} 
We consider several variants of pmc in order to understand the effects of the various components on the method's performance. 
The variants considered are: a serial version with neighborhood cores exploited (pmc), 
the same version but without exploiting neighborhood cores (pmc no neigh cores), and  
a version that uses only the k-core pruning steps and searches vertices in their native 
order, the order in which they were read from disk, rather than degeneracy order (pmc-native).


We compare these three variants of pmc with three state of the art maximum clique finders. These include the recent method FMC  (for fast maximum clique) from \cite{patt2012cliques}, the method MaxCliqueDyn~\cite{konc2007improved} which dynamically adapts a greedy color sort, and a recent implementation of the Bron-Kerbosch (BK) algorithm in the \texttt{igraph} package~\cite{igraph}.



From the performance profile plots, for these types of networks, we find  
little difference between using the neighborhood cores and not using them, and somewhat
more pronounced difference between using degeneracy ordering versus native ordering.
Further, we find a big difference between our most optimized method (pmc)  and 
the alternative algorithms.  Compared to the BK algorithm, we are over 1000 times faster for some problems, and we solve all of the instances. Compared to the FMC algorithm, we are about 50 times faster. This illustrates that our algorithm uses properties of the social and information networks to hone in on the largest clique quickly.

\textbf{Figures~\ref{fig:pp-serial}~and~\ref{fig:pp-16}, DIMACS networks.} These plots show
results on the 32 hard instances of DIMACS problems in the serial case (a) and in the parallel
case (b). It can be seen that, in the serial case, the neighborhood cores greatly help reduce the work in the majority of cases.  In a few cases, they resulted in a large increase in work (the point furthest to the right in the serial figure).  All of the work involved in computing these cores is parallelized, and we observe that, in parallel, using them is never any worse than about $2^{0.5} \approx 144\%$  the speed of the fastest method.  

In summary, we observe that a) our parallelization strategy is effective,
b) our algorithm outperforms existing algorithms dramatically, and
c) neighborhood core bounds help with challenging problems.
We recommend using neighborhood cores as they help the algorithm terminate faster with 
challenging problems and almost never take more than twice the time.

\begin{table}[t!]
\caption{For each temporal network, we list the number of temporal edges, the number of vertices and edges in the reachability graph, the size $\omega$ of the temporal strong component and the runtime of our algorithm.}
%
\label{table:tscc}
\centering
\scriptsize\medskip
\begin{tabularx}{0.85\linewidth}{@{}rXXXHHHXcHH@{}}
\toprule 
\textbf{graph} & $|E_{T}|$ & $|V_{R}|$ & $|E_{R}|$ & $K$ & $\tilde{\omega}$ & $\mu$ & $\omega$ & Time \rlap{(s.)} & $\tau_{heu}$ & $\tau_{\mu}$\\ 
\midrule 
 \textsc{infect}-\textsc{dublin} & 415k & 11k & 176k & 84 & 84 & 1 & 84 & $<$.01 & 1 & 3.38\\ 
 \textsc{infect}-\textsc{hyper} & 20k & 113 & 6.2k & 106 & 106 & 1 & 106 & $<$.01 & 1 & 4.45\\ 
 \midrule
 \textsc{fb}-\textsc{messages} & 61k & 1.9k & 532k & 707 & 707 & 3 & 707 & 0.05 & 1 & 3.95\\ 
 \textsc{reality} & 52k & 6.8k & 4.7M & 1236 & 1236 & 1 & 1236 & 0.19 & 1 & 4.8\\ 
 \midrule
 \textsc{retweet} & 61k & 18k & 66k & 175 & 165 & 1 & 166 & 0.02 & 0.48 & 1.41\\ 
 \textsc{twitter}-\textsc{cop} & 45k & 8.6k & 474k & 583 & 581 & 1 & 581 & 0.22 & 0.55 & 1.44\\ 
\midrule 
\end{tabularx}
\end{table}

\section{Maximum Clique Applications} 

Although finding the maximum clique is generally NP-hard, our procedure is effective on many real-world networks and produces nearly linear runtimes. In this section, we consider how well our method works as a subroutine for two applied problem: finding the largest temporal strong component and finding a compress-friendly order of a network.

\subsection{Temporal strong components}
\label{sec:tscc}

Temporal strong components were recently proposed by Bhadra et al.~and Nicosia et al.~to extend the idea of a strong component in a network to a temporal graph~\cite{bhadra2003complexity,nicosia:023101}. Let $V$ be a set of vertices, and $E_T \subseteq V \times V \times \RR^{+}$ be the set of temporal edges between vertices in $V$. Each edge $(u,v,t)$ has a unique time $t \in \RR^{+}$. For such a temporal network, a path represents a sequence of edges that must be traversed in increasing order of edge times.  That is, if each edge represents a contact between two entities, then a path is a feasible route for information. Temporal paths are inherently asymmetric because of the directionality of time. Two vertices $(u,w)$ are strongly connected if there exists a temporal path $\mathcal{P}$ from $u$ to $w$ and from $w$ to $u$.  A temporal strongly connected component (tSCC) is defined as a maximal set of vertices $C \subseteq V$ such that any pair of vertices in $C$ are strongly connected. Note that this is exactly the same definition as a strong component where we replaced the notion of a path with a temporal path.

As previously mentioned, checking if a graph has a $k$-node temporal SCC is NP-complete~\cite{bhadra2003complexity,nicosia:023101}.  Nonetheless, we can compute the largest such strong component using a maximum clique algorithm.  Let us briefly explain how.  The first step is to transform the temporal graph into what is called a strong-reachability graph.  For each pair of vertices in $V$, we place an edge in the strong reachability graph if there is a temporal path between them.  This is easy to do by using a method by~\cite{Pan-2011-paths}. With this reachability graph, the second step of the computation is to remove any non-reciprocated edges and then find a maximum clique.  That maximum clique is the largest set of nodes where all pairwise temporal paths exist, and hence, is the largest temporal strong component~\cite{nicosia:023101}.

\def \lentab{2.5in}

\begin{figure}[t!]
\centering\subfigure[Reachability (retweet)]
    	{\label{fig:reach-retweets}\includegraphics[width=0.5\linewidth]{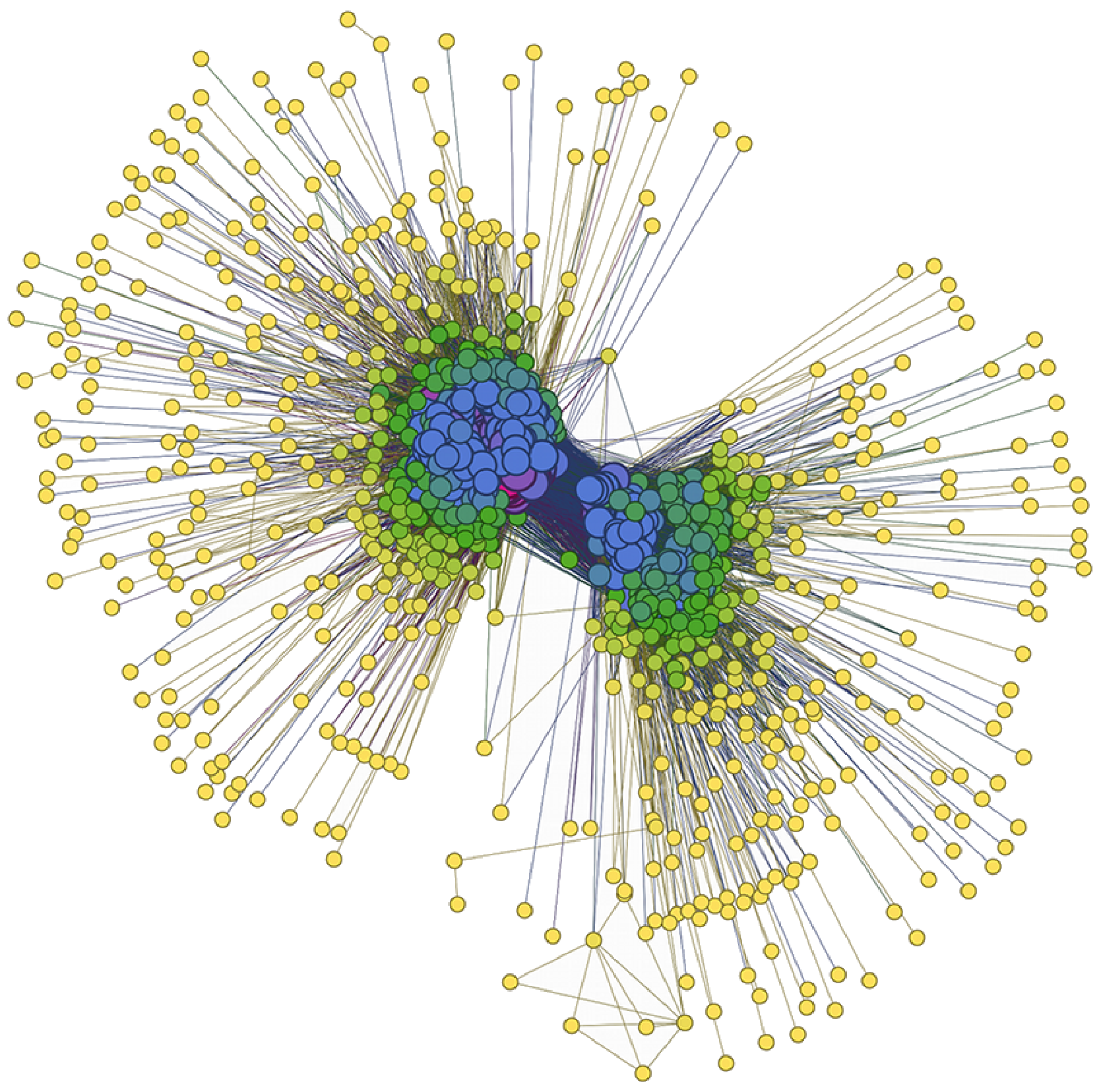}}%
    \subfigure[Temporal SCC (retweet)]
    	{\label{fig:tscc-retweets}\includegraphics[width=0.5\linewidth]{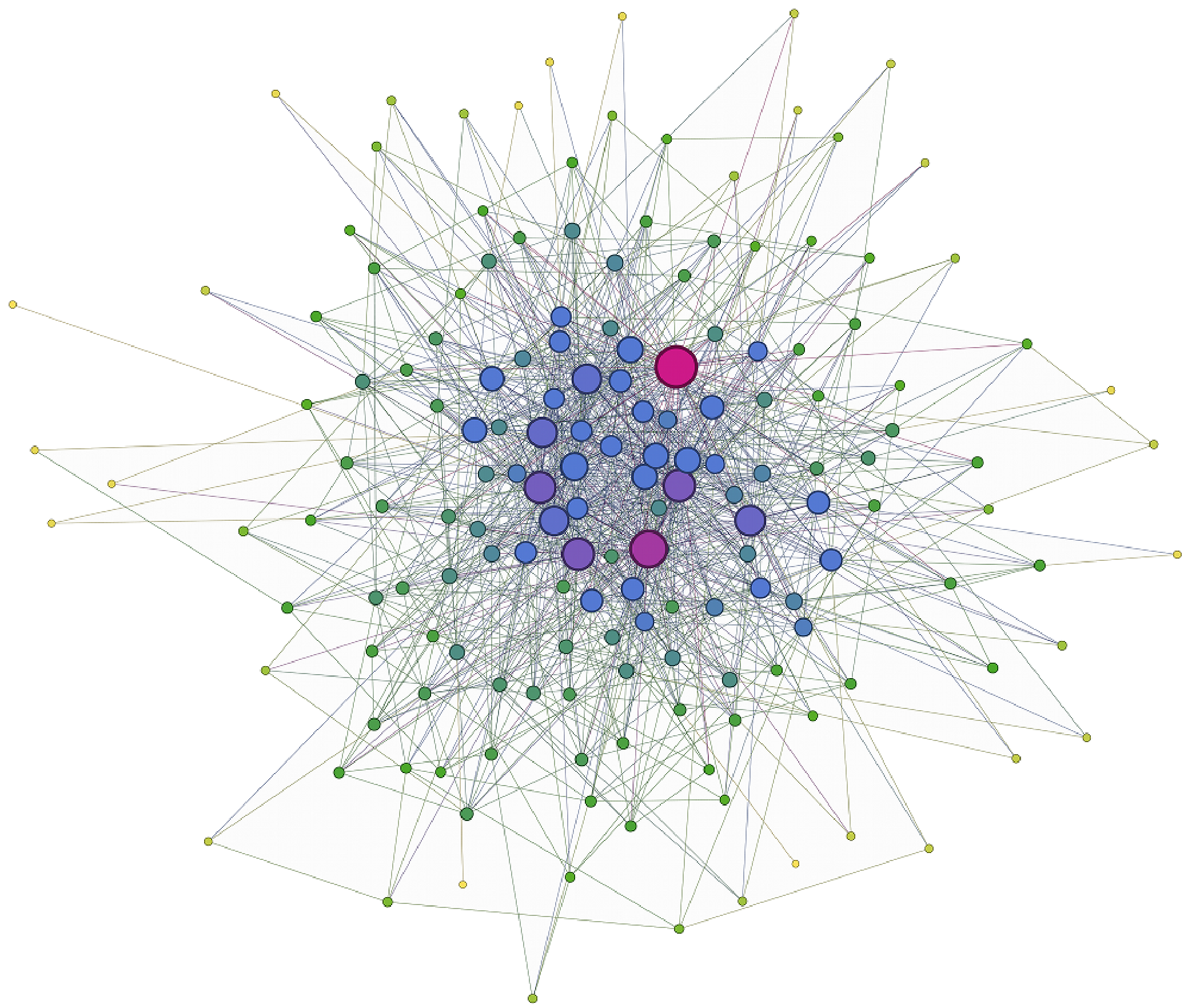}}	
  \caption{In order to compute the largest temporal strong component, we first compute the strong reachability network (a). These networks are rather dense and often reveal clear community structure. Here we see clear communities for the political left and right.  We find that the largest temporal strong component in the retweet network (b) consists of 166 twitter users classified as politically ``right'' according to the original data with only a single exception. }
  \label{fig:tscc-vis}
\end{figure}

\textbf{Data.} We study three types of temporal networks. In each, the nodes represent people. \emph{Contact networks}:
The edges are face-to-face contacts (infect-dublin, infect-hyper\cite{isella2011s}). See ref.~\cite{infect} for more details about these data. \emph{Interaction networks}:
In fb-messages, the edges represent private messages~\cite{opsahl2009clustering}
and in the reality network, the edges represent calls~\cite{eagle2006reality}.
We also investigate a cellular telephone call network where the edges are calls (reality~\cite{eagle2006reality}). \emph{Retweet networks}:
Here, the edges are retweets. We analyzed a network of political retweets centered around the November 2010 election in the US (retweet~\cite{conover2011political}).
A similar dataset is a retweet and mentions network from the UN conference held in Copenhagen. The data was collected over a two week period (twitter-copen~\cite{ahmed2010time}).


\textbf{Results and analysis}.
Figure~\ref{fig:tscc-vis} shows the reachability and largest temporal strong component from a retweet network about politics.  It took the maximum clique finder less than a second to identify this clique.  We summarize the remaining experiments on the temporal strong components in Table~\ref{table:tscc}.  For all of these networks, we were able to identify the largest temporal strong component in less than a second after we computed the reachability network. There are two reasons for this performance.  First, in all of the networks except for the interaction networks, the largest clique is the set of vertices with highest core numbers.  Second, our heuristic computes the largest clique in all of these networks, and we are quickly able to reduce the remaining search space when it isn't the largest $k$-core as well.

We observe several interesting properties in these temporal strong components. In the two contact networks (infect-hyper and infect-dublin), both of the largest strong components had about 100 vertices, despite the drastically different sizes of the initial dataset. We suspect this is a consequence of the data collection methodology since the infect-dublin data were collected over months whereas the infect-hyper data were collected over days.  In the interaction networks, the components contain a significant fraction of the total vertices, roughly 20-30\%.  In the retweet networks, the components are a much smaller fraction of the vertices.  
Given the strong communication pattern between the groups, the components are good candidates for centers communities in the networks. 

Together, these results show that temporal strong components are a strict requirement on a group of nodes in a network.  For instance, there is a considerable difference in the size of temporal strong components between networks with asymmetry in the relations (retweets) compared with networks with symmetric relationships (fb-forum, fb-messages, and reality).  This finding may be important for those interested in designing seeded viral campaigns on these networks.

\subsection{Ordering for network compression}
\label{sec:compress}

In this application, we consider using the maximum cliques of a network to produce an ordering of the vertices that should be useful for reducing the storage space of the network structure. Compression has two important benefits, first, it reduces the amount of IO traffic involved in using the graph; second, good compression schemes may reduce the amount of work involved in running an algorithm on the graph~\cite{karande2009-compressed-graph-algorithms,kang2011beyond}. State-of-the-art network compression techniques heavily exploit locality of links within the adjacency list representation of a graph to reduce the number of bits required to store each edge~\cite{boldi2004-webgraph,boldi2005-codes,boldi2011layered}. Cliques are the densest local feature of a graph and in this application, we order the vertices of a network such that every vertex is in a large clique. This ensures that there are many local edges within the graph. We then evaluate how well the \emph{bvgraph}~\cite{boldi2004-webgraph,boldi2005-codes} compression method reduces the graph size using this ordering. 

The specific ordering we use is the result of the following process. Given a graph $G$, we find a maximum clique $C$ in $G$, remove $C$ from $G$, and repeat the process until all vertices are removed. To improve the runtime, we ran our heuristic method to find large cliques. We then order the vertices according to the cliques, $C_1$, $C_2$, $\ldots$, $C_K$, where $K$ denotes the number of iterations needed. Internally within each $C_i$ we order the vertices by their degrees.  We then permute the graph to use this ordering and use the \emph{bvgraph} compression scheme with all default settings to compress the networks. Table~\ref{tab:pmc-compression} shows the results we get on two Facebook networks. We compare the compression obtained by reporting the size of each graph in bytes after compressing. We evaluate three orderings of the vertices: the native order, the Layered Label Propagation (LLP) order proposed to help improve compression with the bvgraph algorithm~\cite{boldi2011layered}, and our clique-based order computed using PMC. We find that our ordering results in better compression than using the native ordering of the data and it is comparable to the LLP order although slightly worse. Previous research found that identifying and compressing large bicliques with a linear number of edges helped to improve upon methods that use the adjacency list~\cite{buehrer2008-webgraph}. Given the success of this simple ordering, we plan to evaluate these more complicated schemes next.

\begin{table}
\caption{Size in bytes required to store two Facebook graphs using the bvgraph compression scheme in three different orders. } \medskip
\fontsize{8}{9}\selectfont
\label{tab:pmc-compression}
\begin{tabularx}{\linewidth}{@{}lXXXXX@{}}
\toprule
Graph & Vertices & Edges & Native & LLP & PMC \\
\midrule
fb-Penn & 42k & 1.4M & 4237507 & 2740801 & 3104286\\
fb-Texas & 36k & 1.6M & 4605427 & 3232909 & 3508224 \\
\bottomrule
\end{tabularx}
\end{table}

\vspace{3mm}
\section{Related work}
\label{sec:related}

A related problem to maximum clique finding is maximal clique enumeration: identifying {\em all} the maximal cliques in $G$. This problem tends to get more attention in data mining literature~\cite{Xie-2010-max-clique,Cheng-2012-maximal}.  For instance maximal cliques in social networks are distributed according to a power-law~\cite{Du-2009-communication}. There is a considerable body of recent work on this problem~\cite{tomita2006worst,eppstein2011listing,eppstein2010listing,schmidt2009scalable,cheng2011ACM}. In particular, Du et al.~\cite{Du-2009-maximal-cliques}~take advantage of the properties of social and information networks in order to enumerate all maximal cliques faster. In comparison, we wish to highlight how fast we can solve the maximum clique problem for these networks and temporal strong components by appropriately applying pruning steps and bounds. 

Pardalos and Xue \cite{citeulike:4058448} provide a good review of exact algorithms for maximum clique that existed prior to 1994. Notable methods proposed later include: among others, the works of Bomze et al.~\cite{bomze1999maximum}, \"{O}sterg{\aa}rd \cite{ostergard}, Tomita et al.~\cite{Tomita-2007-branch-and-bound}, and San Segundo et al.~\cite{SanSegundo}. In a very recent work, Prosser~\cite{prosser2012} provides a computational study comparing various exact algorithms for maximum clique.
The vast majority of existing work focuses on sequential maximum clique finders. Recent parallel algorithms include a multithreaded algorithm~\cite{mccreesh2013multi}, albeit without using core numbers to accelerate pruning, and a MapReduce-based method~\cite{jingen2013cliqueMapReduce}.


\section{Conclusions}
\label{sec:conclusions}

We propose a new fast algorithm that finds the maximum clique on billion-edge social networks in minutes.  It exhibits linear runtime scaling over graphs from $1000$ vertices to $100$ million vertices and has good parallelization potential.  We applied the algorithm to compute the largest temporal strong components of a dynamic network, which involves finding the largest clique in a static reachability graph, and to obtain an ordering friendly for graph compression.  Our hope is that maximum clique will now become a standard network analysis measure. Towards that end, we make our software package available for others to use: \newline
\centerline{\small
\url{http://www.ryanrossi.com/pmc}
}

\fontsize{7.9}{9.0}\selectfont

\bibliographystyle{abbrv}
\bibliography{rossi,gleich}




\end{document}